# Return and Volatility Forecasting Using On-Chain Flows in Cryptocurrency Markets

*Yeguang Chi[1], Qionghua (Ruihua) Chu[2], Wenyan Hao[3]*
*June 2025*

**Abstract**

We empirically examine the intraday return- and volatility-forecasting power of on-chain flow data for Bitcoin (BTC), Ethereum (ETH), and Tether (USDT). We find ETH net inflows to strongly predict ETH returns and volatility in the 2017-2023 period. Our intraday frequencies are 1–6 hours. We find that differing significantly from forecasting patterns for BTC, ETH net inflows negatively predict ETH returns and volatility. First, we find that USDT flowing out of investors' wallets and into cryptocurrency exchanges, namely, USDT net inflows into the exchanges, positively predicts BTC and ETH returns at multiple intervals and negatively predicts ETH volatility at various intervals and BTC volatility at the 6-hour interval. Second, we find that ETH net inflows negatively predict ETH returns and volatility for all intraday intervals. Third, BTC net inflows generally lack predictive power for BTC returns (except at 4 hours) but are negatively associated with volatility across all intraday intervals. We illustrate our findings on return forecasting via case studies. Moreover, we develop option strategies to assess profits and losses on ETH investments based on ETH net inflows. Our findings contribute to the growing literature on on-chain activity and its asset pricing implications, offering economically relevant insights for intraday portfolio management in cryptocurrency markets.

.

---

[1] University of Auckland Business School, 12 Grafton Road Sir Owen G Glenn Building, Auckland, New Zealand 1010. Email: y.chi@auckland.ac.nz.
[2] Nanyang Technological University, Nanyang Business School, 52 Nanyang Ave, Singapore 639798. Email: chuq0002@e.ntu.edu.sg.
[3] University of Leicester, Department of Mathematics, University Road, Leicester, United Kingdom, LE1 7RH. Email: wenyan881224@gmail.com.

# 1. Introduction

As cryptocurrencies gain prominence in finance, understanding the return-forecasting power of on-chain activities becomes crucial. Despite their unique characteristics and regulatory challenges, cryptocurrencies are increasingly popular in asset management, making it essential to explore whether and how their returns and volatility can be predicted.

We empirically investigate the intraday forecasting power of net inflows[1] for Bitcoin (BTC), Ethereum (ETH), and Tether (USDT) on BTC and ETH returns and volatility. We find ETH net inflows to strongly and negatively predict ETH returns and volatility. USDT, the stablecoin, is meaningful for our empirical analysis to predict BTC and ETH dynamics, as USDT serves as an indicator of buy-side liquidity for cryptocurrency trading (Dow Jones, 2025). Our single-variable models use net inflows (for example, ETH net inflows for ETH returns) as the independent variable, while the double-variable models add BTC and ETH current period returns as independent variables. To ensure robustness, we perform in-sample and out-of-sample regression tests.

Our theoretical rationale for why USDT, BTC, and ETH net inflows should predict BTC and ETH returns and volatility is as follows. Given that BTC and ETH supplies are limited (halving for BTC and burn mechanism for ETH), demand indicated by net inflows influence prices and returns. On volatility, higher net inflows provide greater liquidity and lower transaction costs to reduce volatility, while the higher net outflows drain liquidity and aggravate price swings to increase volatility.

Through our empirical studies, we contribute to several strands of the finance literature. First, we find that USDT net inflows into exchanges predict higher BTC and ETH returns (every 1 and 2 hours) and lower volatility (6-hour intervals for both BTC and ETH and 1-hour intervals for ETH).

Second, our empirical findings focus on ETH return- and volatility-forecasting power. In comparison, studies so far cover mainly anonymous on-chain features of BTC (Filtz et al., 2017; Fleder et al., 2015; Meiklejohn, 2013; Reid & Harrigan, 2013), factors affecting BTC returns and volatility (Alexander et al., 2023; Ante & Fiedler, 2021; Balcilar et al., 2017; Hau et al., 2021; Hoang & Baur, 2020, 2022; Koutmos, 2018; Kristoufek, 2015; Panagiotidis et al., 2018), spot trading volume and liquidity on less widespread cryptocurrencies such as Dash, Litecoin, and NEM (Bouri et al., 2019), and positive relationship between volatility of liquidity and returns for large capitalization cryptocurrencies other than BTC (Leirvik, 2021). Drawing parallels to long-term return predictability in traditional assets such as equities (Ang & Bekaert, 2007; Boudoukh et al., 2019), we investigate the intraday predictability of returns and volatility. Whereas features of transactions could affect the realized volatility of stocks (Chan & Fong, 2006), we explore how net inflows of transactions could forecast ETH returns and volatility. While Hoang and Baur (2020, 2022) confirmed BTC net inflows to forecast negative BTC returns and volatility, we find that ETH net inflows and current period ETH returns and volatility demonstrate robust intraday forecasting power for future ETH returns and volatility respectively for all intraday intervals, namely, 1, 2, 3, 4, and 6 hours.

---

[1] Net inflows are calculated as the difference between inflows (from wallets to exchanges) and outflows (from exchanges to wallets).

Lastly, we conduct a similar empirical analysis on BTC and find out that apart from the interval of 4 hours which renders a positive relationship, BTC net inflows lack predictive power on BTC returns. As for volatility, BTC net inflows negatively predict the volatility of BTC for all intraday intervals.

To view our results in specific contexts, we conduct case studies to confirm our empirical findings that USDT net inflows positively predict BTC and ETH returns, ETH net inflows negatively predict ETH returns, and BTC net inflows generally lacks predictive power on BTC returns, except at 4-hour intervals.

For real-world applications, we indicate portfolio management and trading implications and assess profitability of various option strategies to make gains and minimise losses on ETH investments based on ETH net inflows.

The remainder of our paper is organized as follows. Section 2 describes the data and methodology. Section 3 discusses and analyzes our main results and conducts robustness checks for concerns and alternative explanations. Section 4 demonstrates asset management and trading applications and how option strategies could be developed to make profits. Section 5 concludes the paper findings and extends research insights.

## 2. Data and Methodology

### 2.1. Data Source and Robustness Check

As indicated by Alexander and Dakos (2020) for the importance of data quality in cryptocurrency research, we use high-quality cryptocurrency data from a range of sources[2]. Our sample period is from December 16, 2017 to January 20, 2023.

For a comprehensive analysis, as per Griffin and Shams (2020), as BTC might be influenced by USDT movements because of supply, we also conduct empirical studies on how USDT net inflows might affect BTC and ETH returns.

As compared to the literature by Hoang and Baur (2020, 2022), with BTC as a starting point, our empirical research focuses on ETH and confirms our hypotheses that the net inflow of ETH indicates a negative return because of the movement to sell, with robustness checks performed.

For robustness check, we conduct an intraday return- and volatility-forecasting power regression analysis for the interval 1, 2, 3, 4, and 6 hours to cover multiple frequencies.

**2.2. Intraday Return-Forecasting for ETH and BTC Methodology**

We investigate the return-forecasting power of net inflows on returns based on the following regressions, with (1) for single-variable and (2) for double-variable models for the forecasting interval of 1 hour. For the intervals of 2, 3, 4, and 6 hours, we apply similar regressions, with t+1 updated to t+2, t+3, t+4, and t+6 accordingly. Net inflows are in the unit of US$1 million, and returns are expressed as decimals. We report results in Tables 2 and 3. The magnitude of coefficients is less important than signs to indicate the directions of statistically significant relationships to provide economic implications.

$$R_{t+1} = \beta_0 + \beta_1 \times \text{Net Inflow}_t + e_{t+1} \text{ and} \qquad (1)$$

$$R_{t+1} = \beta_0 + \beta_1 \times \text{Net Inflow}_t + \beta_2 \times R_t + e_{t+1}, \qquad (2)$$

where R and e are returns and residuals; t and t+1 stand for current and next 1-hour periods. $\beta_0$ is intercept; $\beta_1$ is coefficient on net inflow; $\beta_2$ is control variable (return at t); standard errors in parentheses.

**2.3. Intraday Volatility-Forecasting for ETH and BTC Methodology**

We investigate the volatility-forecasting power of net inflows on volatility based on regressions (3) and (4), with (3) for single-variable and (4) for double-variable models for the forecasting interval

---
[2] The list of exchanges for data sources are BiBox, BigONE, Binance, Bitfinex, Bitget, Bithumb, BitMEX, Bitstamp, Bittrex, Bybit, CheckSig, Cobinhood, Coinbase, Coincheck, CoinEx, Crypto.com, Deribit, Gate.io, Gemini, HitBTC,

of 1 hour. For other intraday intervals, we apply similar regressions, with t+1 updated to t+2, t+3, t+4, and t+6 accordingly. Net inflows are in the unit of US$1 million, and volatility is expressed as decimals. We report results in Tables 4 and 5. Similar to Section 2.2., signs of coefficients are more important than the magnitude.

$$V_{t+1} = \beta_0 + \beta_1 \times \text{Net Inflow}_t + e_{t+1} \text{ and} \tag{3}$$

$$V_{t+1} = \beta_0 + \beta_1 \times \text{Net Inflow}_t + \beta_2 \times V_t + e_{t+1}, \tag{4}$$

where V and e are volatility and residuals, with t and t+1 standing for current and next period. $\beta_0$ is intercept; $\beta_1$ is coefficient on net inflow; $\beta_2$ is control variable (return at t); standard errors in parentheses.

## 2.4. Case Studies on Extreme Events and Option Strategies Methodology

For computations, we apply the following formulas, where $P_{call}$, $P_{option}$, $P_{index,current\ quote\ time}$, $P_{index,later\ quote\ time}$, $R_{sell\ call\ option}$, $R_{buy\ call\ option}$, $P_{call,current\ quote\ time}$, $P_{call,later\ quote\ time}$, $\delta$, $PnL_{option}$, $PnL_{underlying}$, $PnL_{portfolio}$, and $R_{portfolio}$ stand for price of call option, unit price of option benchmarked to ETH, index price – underlying price – at the current quote time, index price at a later quote time, return of selling call options, return of buying call options, call option price at the current quote time, call option price at the later quote time, delta unit of underlying, PnL of options, PnL of underlying, and PnL and returns of portfolio for selling ETH options. We assume $10,000 as the initial capital for strategies. We report results in Table 6.

$$P_{call} = P_{option} \times P_{index}$$

$$PnL_{option} = (P_{call,current\ quote\ time} - P_{call,later\ quote\ time})$$

$$R_{sell\ call\ option} = \frac{PnL_{option}}{P_{call,current\ quote\ time}}$$

$$R_{buy\ call\ option} = -R_{sell\ call\ option}$$

$$R_{underlying} = \delta \times R_{buy\ call\ option}$$

$$PnL_{underlying} = (P_{index,later\ quote\ time} - P_{index,current\ quote\ time}) \times \delta$$

$$PnL_{portfolio} = PnL_{option} + PnL_{underlying}$$

$$R_{portfolio} = \frac{PnL_{portfolio}}{10,000}$$

## 2.5. Profitability of Option Strategies in Various Scenarios Methodology

In Table 7, we report profitability of sell call top versus bottom net return percentile based on profits and losses comparisons. Besides, in Table 7, we report profitability of sell call top versus bottom buy call net return percentile based on profits and losses comparisons. Moreover, in Table 8, we report profitability of sell call options top percentiles in breakeven scenario. Top and bottom percentiles are based on net inflows of ETH. For Tables 7, 8, and 9, we have tested Implied Volatilities (IV) less than 1, 2, 3, and whole sample (4.1 for top percentiles and 4.5 for bottom percentiles) and greater than or equal to 1, 2 and 3. For Out of the Money (OTM) range, we have tested less than 1%, between 1% and 2%, 2% and 3%, 3% and 4%, 4% and 5%, 1% to 3%, 3% to 5%, 5% to 10%, 10% to 15%, and 15% to 20%. We have also tested the combinations of IV and OTM ranges for the categories, such as IV less than 1 and OTM less than 1%. We have kept the most evident results in the three tables, and other combinations have similar results.

We compute the variables in Tables 7 to 9 by measuring the ETH call options with the same strike price, time to maturity, and quote time. We apply the following formulas to compute the variables.

For both buy and sell call options, we adopt the following formula, with call price representing price of the call options with the same strike price, time to maturity, and quote time.

$$P_{call} = P_{option} \times P_{index}$$

To decide if the trading strategies are profitable, we apply the following formulas.

$$\text{Win Rate} = \frac{\text{Total Number of Win Trades}}{\text{Total Number of Trades}}$$

If Net Portfolio Return > 0, the trade is defined as a "win" trade; otherwise, it is defined as a losing trade. "WtL" stands for "Win-to-Loss" ratio.

$$\text{WtL} = \frac{\text{Total Number of Win Trades}}{\text{Total Number of Loss Trades}}$$

PnL stands for profits and losses. Besides, $PnL_{portfolio}$, $PnL_{option}$, and $PnL_{portfolio,net}$ stand for profits and losses for portfolios, options, and net portfolios respectively. Also, $\delta$ stands for option delta.

$$PnL_{portfolio} = PnL_{option} + PnL_{portfolio,net}$$

$$PnL_{portfolio,net} = PnL_{portfolio} - (0.0003 + 0.0005 \times \delta + \frac{0.001}{2} + \text{Slippage}) \times P_{index}$$

$0.0003$, $0.0005 \times \delta$, $\frac{0.001}{2}$, and Slippage represent option premiums of 0.03% of Index Price, perpetual hedge of $0.0005 \times \delta \times$ Underlying, and bid-ask spread of $\frac{0.001}{2} \times$ Index Price, and Slippage of 0 for Tables 6 and 8 as a base scenario, and Slippage of an unknown number to make $PnL_{portfolio,net}$ zero for Table 8.

$Range_{OTM}$, $P_{strike}$, $R_{portfolio,net}$, and δ stand for Out of the Money (OTM) range, strike price, net portfolio return, and option delta.

$$Range_{OTM} = \frac{(P_{strike} - P_{index})}{P_{index}}$$

$$Initial\ Capital = (0.3 + δ) \times P_{index}$$

## 3. Main Results

### 3.1. Key Empirical Contributions

Our empirical analysis provides novel evidence of the predictive power of USDT net inflows in intraday cryptocurrency returns and volatility. Besides, our findings differentiate clearly between BTC and ETH forecasting dynamics and provide nuanced insights into cryptocurrency-specific behaviors. In Table 1 Panels A and B, we show our summary findings and contributions.

### 3.2. USDT Net Inflows Positively Predict ETH and BTC Returns

*3.2.1. Robustness of Results and Economic Interpretation*

In Panel A of Tables 2 and 3, we demonstrate outcomes of predictive regression of both single-variable and double-variable models. We have tested 5 intraday intervals for robustness. We find that USDT net inflows positively predict both ETH and BTC returns, especially for intervals of 1 and 2 hours for both models. USDT net inflows are dry powder[3] representing available buy-side liquidity and result in ETH and BTC return increase.

However, we are aware that the forecasting relationship of USDT net inflows on BTC and ETH returns is not statistically significant for either models for the intervals of 3, 4, or 6 hours.

*3.2.2. Magnitude of Impact*

We interpret magnitude of impact from $β_1$ values in the double-variable models, as it considers a control and provides more robust results. To quantify impact, as per Panel A of Tables 2 and 3,

---

[3] Dry powder indicates readily deployable capital for investing in the cryptocurrency market.

US$ 100 million of USDT net inflows predict 0.11% of ETH return and 0.065% of BTC return in the next hour.

### 3.3. ETH Net Inflows Negatively Predict ETH Returns

*3.3.1. Robustness of Results and Economic Interpretation*

Based on results in Table 2, Panel B for regressions (1) and (2) on ETH net inflows to forecast ETH returns, for both single- and double-variable models, we find statistically significant intraday negative relationships between ETH net inflows and returns across all intraday intervals of 1, 2, 3, 4, and 6 hours. We have examined all intraday intervals for robustness.

For ETH, the net inflows of ETH indicate movement to sell ETH, suggesting a bearish market. Hence, returns on ETH will be negative. ETH net inflows indicate sell-side pressure and lead to lower ETH return.

As for net outflows of ETH, we find no consistent relationship on returns. While net outflows of ETH might be the result of buying ETH to suggest positive ETH returns, the ETH might not be used to consistently push up ETH prices to result in positive returns. Thus, we do not observe empirical evidence on the relationship between ETH net outflows and returns.

Interestingly, while net inflows to exchanges exhibit strong predictive power for subsequent returns and volatility, net outflows do not display a symmetric or consistent relationship. One plausible economic explanation is that outflows from exchanges to wallets can serve multiple purposes beyond imminent trading activity, such as long-term custody, cold storage security, or strategic reserve management, which do not necessarily imply immediate buy-side pressure. In contrast, inflows to exchanges are more directly indicative of sell intent, as cryptocurrencies are typically transferred to trading platforms when users anticipate converting them into fiat or other assets. This inherent asymmetry between the informational content of inflows versus outflows further validates our focus on net inflow measures in forecasting models.

*3.3.2. Magnitude of Impact*

As the double-variable models consider a control and provides more robust results, we interpret magnitude of impact from $\beta_1$ values in the double-variable models. To quantify impact, as per Table 2, Panel B, US$ 1 million of ETH net inflows predict -1.70% of ETH return in the next hour.

### 3.4. BTC Net Inflows Positively Predict BTC Returns

*3.4.1. Robustness of Results and Economic Interpretation*

As seen from Table 3, Pabel B, based on regressions (1) and (2), for both single- and double-variable models, at the interval of 4 hours, BTC net inflows positively predict BTC returns, but not for other intervals. For robustness, we have examined all intraday intervals. BTC net inflows provide mostly weak predictive power on BTC returns, suggesting different investor behaviors. Potential lag in order execution and broader network liquidity dynamics are possible reasons for why only the 4-hour intervals matter. On economics and market structure, this implies that significant buying is more likely to be observed 4 hours after BTC net inflows occur to drive BTC prices and form positive returns. Overall, for BTC, we observe no consistent intraday positive relationship exists between its net inflows and returns.

*3.4.2. Magnitude of Impact*

Since the double-variable models consider a control and provides more robust results, we interpret magnitude of impact from $β_1$ values in the double-variable models. To quantify impact, as per Table 3, Panel B, US$ 1 million of BTC net inflows predict 0.99% of BTC return in the next 4-hour interval.

**3.5. USDT Net Inflows Negatively Predict Volatility of ETH and BTC**

*3.5.1. Robustness of Results and Economic Interpretation*

We investigate the volatility-forecasting power of net inflows on volatility (V) based on regressions (3) and (4). We have examined all intraday intervals for robustness. In Panel A of Tables 4 and 5, we report outcomes of predictive regression of single and double independent variables models respectively. We observe that at the 6-hour interval, USDT net inflows negatively predict volatility for both ETH and BTC for both single- and double-variable models. Besides, at the 1-hour interval, USDT net inflows negatively predict ETH volatility for the single-variable models. For other intraday intervals, USDT net inflows do not significantly forecast volatility for either ETH or BTC.[4] This implies that ETH negative volatility tends to be observed 1 and 6 hours following USDT net inflows and BTC negative volatility tend to be observed 6 hours following USDT net inflows. USDT net inflows are dry powder and result in ETH and BTC volatility dampening.

---

[4] While our focus is on intraday prediction, for completeness, Appendix A table Panels A and B report volatility-forecasting results at daily and weekly intervals for USDT net inflows on ETH and BTC volatility.

While USDT produces some statistically significant results to forecast volatility of ETH, the results are not consistent across sample periods tested. Besides, some results are opposite to our hypotheses. Hence, we do not find consistent statistically significant relationships of USDT net inflows to forecast volatility for ETH.

*3.5.2. Magnitude of Impact*

We interpret magnitude of impact from $\beta_1$ values in the double-variable models, because it considers a control and provides more robust results. To quantify impact, as per Panel A of Tables 4 and 5, US$ 1 million of USDT net inflows predict -1.9 bps of ETH and BTC volatility in the next 6-hour interval.

**3.6. ETH Net Inflows Negatively Predict Volatility of ETH**

*3.6.1. Robustness of Results and Economic Interpretation*

For the same return-forecasting investigation based on regressions (3) and (4), in Table 4, Panel B, we report outcomes of predictive regression of single and double independent variables models respectively. For robustness, we have examined all 5 intraday intervals.

For the single-variable models, a negative relationship exists between their ETH net inflows and volatility across all intraday intervals of 1, 2, 3, 4, and 6 hours. As for the double-variable models, a negative relationship exists between ETH net inflows and volatility for all intraday intervals except the 1-hour interval.[5]

While the 1-hour intervals only have predictive power based on the single-variable models, ETH net inflows negatively predict ETH volatility for all intraday intervals. ETH net inflows indicate sell-side pressure and lead to lower ETH volatility.

Similar to Section 3.2., ETH net inflows indicate selling pressure, suggesting bearish market sentiment. While traditionally greater trading activity is associated with heightened volatility, our findings suggest that in the cryptocurrency market, large net inflows to exchanges—signaling selling pressure—may lead to immediate price declines followed by a stabilization phase. In highly liquid periods after large sell-offs, the lower price base reduces the magnitude of subsequent price swings, resulting in lower realized volatility. This is particularly relevant for cryptocurrencies where rapid liquidation events (for example, margin calls, stop-loss cascades) can compress volatility in the aftermath. Therefore, our empirical observation that ETH and BTC net inflows predict lower volatility is economically intuitive when considering the market's tendency toward post-liquidation stability following intense selling.

---

[5] While our focus is on intraday prediction, for completeness, Appendix A table Panel C reports volatility-forecasting results at daily and weekly intervals for ETH net inflows on ETH volatility.

*3.6.2. Magnitude of Impact*

As it considers a control and provides more robust results, we interpret magnitude of impact from $\beta_1$ values in the double-variable models. To quantify impact, as per Table 4, Panel B, US$ 1 million of ETH net inflows predict -0.37% of ETH volatility in the next 2-hour interval.

### 3.7. BTC Net Inflows Negatively Predict Volatility of BTC

*3.7.1. Robustness of Results and Economic Interpretation*

In Table 5, Panel B, on the single-variable models, a negative relationship exists between BTC net inflows and volatility across all intraday intervals of 1, 2, 3, 4, and 6 hours. For the double-variable models, a negative relationship between BTC net inflows and volatility only exists for the 3- and 6-hour intervals. We have examined all 5 intraday intervals for robustness.[6]

While the 1-, 2- and 4-hour intervals only have predictive power based on the single-variable models, BTC net inflows negatively predict BTC volatility for all intraday intervals. BTC net inflows provide relatively good predictive power on BTC volatility. BTC net inflows provide mostly weak predictive power on BTC volatility, suggesting different investor behaviors.

*3.7.2. Magnitude of Impact*

As it considers a control and provides more robust results, we interpret magnitude of impact from $\beta_1$ values in the double-variable models. To quantify impact, as per Table 5, Panel B, US$ 1 million of BTC net inflows predict -2.8% of BTC volatility in the next 3-hour interval.

### 3.8. Case Studies

*3.8.1. Case Studies Extreme Events and Option Strategies*

In Figure 1.a. and Figure 1.b., we demonstrate case studies on USDT and ETH extreme net inflows and 1-hour ETH call option close price on May 11 to May 12, 2022 and January 18 to January 22, 2022. We have identified those two events to be extreme for case studies. In Figure 1.c. and

---

[6] While our focus is on intraday prediction, for completeness, Appendix A table Panel D reports volatility-forecasting results at daily and weekly intervals for BTC net inflows on BTC volatility.

Figure 1.d., we demonstrate case studies on USDT and BTC extreme net inflows and 1-hour BTC call option close price on November 6 to November 10, 2022 and May 9 to May 14, 2022. We have identified those two events to be extreme for case studies. Stablecoin slumped on those days to result in extreme market movements. Besides, those days fall in the extreme events filtered by our criteria. We identify extreme events by the unit of hours for which ETH prices experience extreme net inflows to be within the top 10 net inflows of 2021 and 2022 respectively, i.e., top 99.89$^{th}$ percentile, $1 - \frac{10}{24 \times 365} = 99.89$ percentile.

We confirm our hypothesis that on those days with extreme market movements, for 1-hour frequency, USDT net inflows positively forecast ETH and BTC returns, ETH net inflows negatively forecast ETH returns, and no significant pattern is observed for BTC net inflows to forecast BTC returns.

Take an example. As seen from Figure 1.a., on May 11, 2022, extreme net inflows of ETH, but not USDT, results in extreme drop in 1-hour ETH close price; as for May 12, 2022, extreme net inflows of both ETH and USDT result in extreme drop in 1-hour ETH close price. On the other hand, on both days, ETH net inflows are followed by negative 1-hour ETH returns, as evidenced by the downward ETH price movements one hour later. This further demonstrates that the statistically significant relationship of ETH net inflows to forecast bearish ETH movements and returns, and the relationship for USDT net inflows to forecast bearish ETH movements and returns, while occasionally significant, is not consistent throughout.

For 2-, 3-, 4-, and 6-hour ETH close prices, trends are similar to the 1-hour close price in forecasting returns with ETH net inflows. USDT net inflows show a similar trend only with the 2-hour ETH close price, aside from the 1-hour frequency. These observations confirm a consistent statistically significant relationship between ETH net inflows and bearish returns, while the relationship with USDT inflows is occasionally significant but inconsistent.

For May 11, 2022, extreme net inflows of ETH, but not USDT, results in extreme drop in 1-hour ETH close price. As for May 12, 2022, extreme net inflows of both ETH and USDT result in extreme drop in 1-hour ETH close price. This further demonstrates that the statistically significant relationship of ETH net inflows to forecast bearish ETH movements and returns, and the relationship for USDT net inflows to forecast bearish ETH movements and returns, while occasionally significant, is not consistent throughout. For 2-, 3-, 4-, and 6-hour ETH close price, they present similar trends as 1-hour close price for ETH net inflows to forecast ETH returns, and only 2-hour

ETH close price presents similar trends as 1-hour close price for USDT net inflows to forecast ETH returns.

In Table 6, we report extreme events and corresponding implied volatilities of call strategies. We identify extreme events by the unit of hours for which ETH prices experience extreme net inflows to be within the top 99.89$^{th}$ percentile of 2021 and 2022 respectively. The extreme events are identified as the events with corresponding major market news occurring. Those extreme events occur at 12:00, 12$^{th}$ May 2022, 14:00, 11$^{th}$ May 2022, 10:00, 28$^{th}$ Jul 2021, 06:00, 28$^{th}$ Jul 2021, and 13:00, 20$^{th}$ Jul 2021. We identify returns of options based on call prices at different option quote time for options with the same strike price and maturity. Except for 10:00 28$^{th}$ Jul 2021 and 06:00, 28$^{th}$ Jul 2021, selling ETH call options is a profitable strategy. Hence, we confirm our hypothesis that selling ETH call options is a profitable strategy when net inflow is high, i.e., movement to sell ETH, and ETH market experiences a bearish outlook.

## 4. Portfolio Management and Trading Implications

### 4.1. ETH and BTC Portfolio Management and Trading Strategies

Economically, ETH net inflows to negatively predict the ETH returns is intuitive, as selling pressure will reduce prices and returns. This gives rise to practical implications that portfolio managers and traders should buy ETH after 1, 2, 3, 4, or 6 hours when net inflows for ETH occur to enter the market at relatively lower prices. For BTC, as net inflows at 4 hours predict positive returns, selling could enable portfolio managers and traders to gain relatively higher returns. On volatility, the selling pressure suggested by ETH and BTC net inflows at all intraday intervals (1, 2, 3, 4, and 6 hours) negatively predict volatility is consistent with the post-liquidation stabilization commonly observed in cryptocurrency markets. As for USDT, portfolio managers and traders should sell when net inflows at 1 and 2 hours occur to capture positive returns for BTC and ETH, and expect USDT net inflows to negatively predict ETH volatility in the next 1-hour intervals and negatively predict BTC and ETH volatility for the next 6-hour intervals because of the ensuing lowered prices based on prices already lowered.

### 4.2. Profitability of ETH Option Strategies in Various Scenarios

ETH net inflows consistently forecast ETH returns for both single- and double-variable models. Hence, we develop various option strategies on ETH investments to assess how portfolio managers and trades could make profits and minimise losses based on ETH net inflows.

As seen from Table 7, sell call options generate profitable trades for most of the scenarios in top 10%, 5%, and 1% of ETH net inflows and unprofitable trades for most of the bottom 10%, 5%, and 1% of ETH net inflows. This is consistent with our hypothesis that with ETH net inflows, investors are selling ETH and sell call options should be profitable. As the bottom percentiles of ETH net inflows represent ETH outflows, sell call options should be unprofitable.

As per Table 8, sell call options generate profitable trades for most of the scenarios in top 10%, 5%, and 1% of ETH net inflows and buy call options generate unprofitable trades for most of the bottom 10%, 5%, and 1% of ETH net inflows. This is consistent with our hypothesis that with ETH net inflows, investors are selling ETH and sell call options should be profitable. As the bottom percentiles of ETH net inflows represent ETH outflows, buy call options should be unprofitable.

With ETH net inflows, movements to sell form a bearish market. Selling ETH call options is a derivative way to sell ETH. We form and confirm our hypothesis that selling ETH call options will be profitable, based on results in Table 4, Panel B.

In comparison, while ETH net outflows could result from investors buying ETH, this might not be the only reason. Hence, we hypothesize that neither selling nor buying ETH call options, i.e., selling or buying ETH in essence, will be profitable. Table 4, Panel B results prove our hypothesis.

Based on Table 9, sell call options breakeven win rate and win to loss ratios are found for top 10%, 5%, and 1% of ETH net inflows. As the win rate and win to loss ratios are relatively sensible, this is consistent with our hypothesis that with ETH net inflows, investors are selling ETH and sell call options have reasonable breakeven.

## 5. Conclusion

In our paper, we form novel and impactful findings. Different from prior BTC-focused studies, we find consistent evidence of ETH intraday net inflows negatively predicting ETH returns and volatility for all intervals. For BTC on BTC, while BTC net inflows at the 4-hour interval positively predict BTC returns, BTC net inflows negatively predict the volatility of BTC. As for USDT, its net inflows positively predict ETH and BTC returns for the intervals of 1 and 2 hours, while negatively predict the volatility of ETH and BTC at the 6-hour interval. For robustness, we validate our findings through in-sample and out-of-sample regression analysis. On practical applications, we assess portfolio management and trading strategies and develop various scenarios to find out profitable option strategies on ETH based on ETH net inflows.

For the broader implications of our findings, first, practice-wise, the herding behavior of selling to form ETH net inflows might negatively predict the ETH returns and volatility. While BTC net inflows at the 4-hour interval positively predict BTC returns, the investors' herding behavior of selling to form BTC net inflows negatively predicts the volatility of BTC for 1-, 2-, 3-, 4-, and 6-hour intervals. Second, exchange regulation-wise, regulators might not have to be worried about malevolent trading distortions if they observe the following patterns. For large ETH net inflows, negative ETH returns and volatility could ensue; on large BTC net inflows, positive BTC returns for the 4-hour interval and negative BTC volatility for 1-, 2-, 3-, 4- and 6-hour intervals could follow. Lastly, stablecoin regulation-wise, while regulators might not have to set guidelines to control positive ETH and BTC returns at 1- and 2-hour intervals and negative volatility of ETH and BTC at 6-hour intervals because of USDT net inflows, they might want to be cautious about statistically significant positive or negative ETH and BTC returns and volatility at other intervals and set forth compliance requirements to prevent market manipulations.

As for future research to address the limitations of our findings, we believe it might be valuable to explore firstly, more data periods to cover both intraday and multi-day return- and volatility-forecasting power for BTC and ETH. Moreover, the research could go beyond BTC and ETH to cover other major or emerging cryptocurrencies. Finally, exploring via machine learning techniques might also be interesting.

# Appendix A: Daily and Weekly Net Inflow Forecasting Power on Volatility

In this table, we present regression results on the volatility-forecasting power of net inflows. 'USDT v ETH' and 'ETH v ETH' use USDT and ETH net inflows correspondingly to predict the volatility of ETH; 'USDT v BTC' and 'BTC v BTC' use USDT and BTC net inflows correspondingly to predict the volatility of BTC. Our data sources are a list of exchanges[7]. The first and second rows for each pair represent single- and double-variable models correspondingly. Volatility refers to the volatility of ETH, and net inflows are inflows (wallets to exchanges) minus outflows (exchanges to wallets). $\beta_0$ is the intercept, $\beta_1$ is the net inflow coefficient (of primary interest), and $\beta_2$ is the control variable coefficient (ETH return at time t); standard errors are in parentheses. The sample spans December 16, 2017 to January 20, 2023. We use ***, **, and * to denote significance at the 1%, 5%, and 10% levels (two-sided), correspondingly.

**Panel A. USDT Net Inflows on the Volatility of ETH**

| Variable | $\beta_0$ | $\beta_1$ | $\beta_2$ | $R^2$ Adjusted |
|---|---|---|---|---|
| Daily volatility - USDT v ETH | 0.88*** | $-1.3 \times 10^{-5}$ | | -0.001 |
| | (74.646) | (-0.130) | | |
| | 0.28*** | $-6.5 \times 10^{-6}$ | 0.68*** | 0.478 |
| | (16.368) | (-0.088) | (41.062) | |
| Weekly volatility - USDT v ETH | 0.91*** | $8.4 \times 10^{-5}$ | | 0.002 |
| | (32.029) | (1.237) | | |
| | 0.42*** | $1.2 \times 10^{-4}$** | 0.53*** | 0.275 |
| | (7.739) | (2.132) | (9.898) | |

**Panel B. USDT Net Inflows on the Volatility of BTC**

| Variable | $\beta_0$ | $\beta_1$ | $\beta_2$ | $R^2$ Adjusted |
|---|---|---|---|---|
| Daily volatility - USDT v BTC | 0.69*** | $-1.9 \times 10^{-5}$ | | -0.001 |
| | (67.104) | (-0.211) | | |
| | 0.19*** | $7.5 \times 10^{-6}$ | 0.72*** | 0.519 |
| | (14.474) | (0.122) | (44.558) | |
| Weekly volatility - USDT v BTC | 0.72*** | $6.6 \times 10^{-5}$ | | 0.001 |
| | (29.451) | (1.130) | | |
| | 0.27*** | $1.0 \times 10^{-4}$** | 0.61*** | 0.377 |
| | (6.728) | (2.208) | (12.530) | |

**Panel C. ETH Net Inflows on the Volatility of ETH**

| Variable | $\beta_0$ | $\beta_1$ | $\beta_2$ | $R^2$ Adjusted |
|---|---|---|---|---|
| Daily volatility - ETH v ETH | 0.88*** | -0.44*** | | 0.006 |
| | (75.451) | (-3.607) | | |
| | 0.29*** | -0.029 | 0.67*** | 0.458 |
| | (16.751) | (-0.318) | (39.494) | |
| Weekly volatility - ETH v ETH | 0.92*** | -0.19** | | 0.015 |
| | (33.691) | (-2.238) | | |
| | 0.45*** | -0.069 | 0.52*** | 0.271 |
| | (8.185) | (-0.927) | (9.658) | |

**Panel D. BTC Net Inflows on the Volatility of BTC**

| Variable | $\beta_0$ | $\beta_1$ | $\beta_2$ | $R^2$ Adjusted |
|---|---|---|---|---|
| Daily volatility - BTC v BTC | 0.69*** | -7.7*** | | 0.017 |
| | (68.018) | (-5.756) | | |
| | 0.19*** | 0.40 | 0.72*** | 0.512 |
| | (14.367) | (0.421) | (43.009) | |
| Weekly volatility - BTC v BTC | 0.72*** | -3.9** | | 0.061 |
| | (30.999) | (-4.214) | | |
| | 0.30*** | -1.4* | 0.58*** | 0.372 |
| | (7.338) | (-1.770) | (11.389) | |

---

[7] Our data sources are BiBox, BigONE, Binance, Bitfinex, Bitget, Bithumb, BitMEX, Bitstamp, Bittrex, Bybit, CheckSig, Cobinhood, Coinbase, Coincheck, CoinEx, Crypto.com, Deribit, Gate.io, Gemini, HitBTC, Huobi, Korbit, Kraken, KuCoin, Luno, Nexo, OKX, Poloniex, SwissBorg, and ZB.com.

**Figure 1**

**Case studies on USDT and ETH extreme net inflows and ETH call option close price**

Figure 1 shows the USDT and ETH net inflows and ETH call option close prices in extreme market movements. Our data sources are a list of exchanges[8]. Our sample period is from January 1, 2021 to December 31, 2022. Our case studies cover extreme events on May 11, 2022 and May 12.

**Figure 1.a**. USDT and ETH Net Inflows to Forecast ETH Returns in 1 Hour from May 9 to May 14, 2022

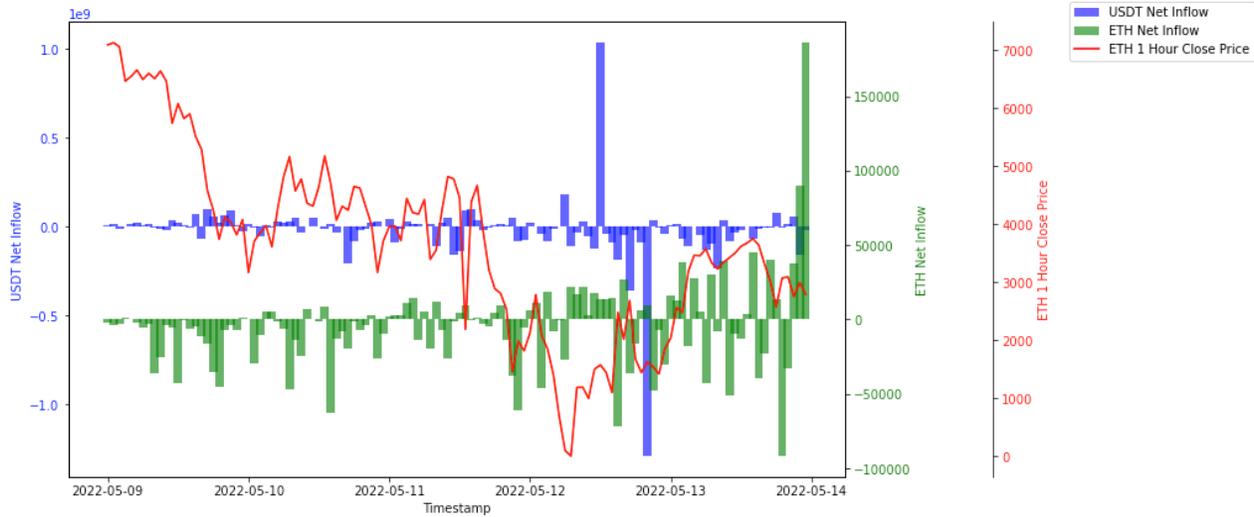

**Figure 1.b**. USDT and ETH Net Inflows to Forecast ETH Returns in 1 Hour from January 18 to January 22, 2022

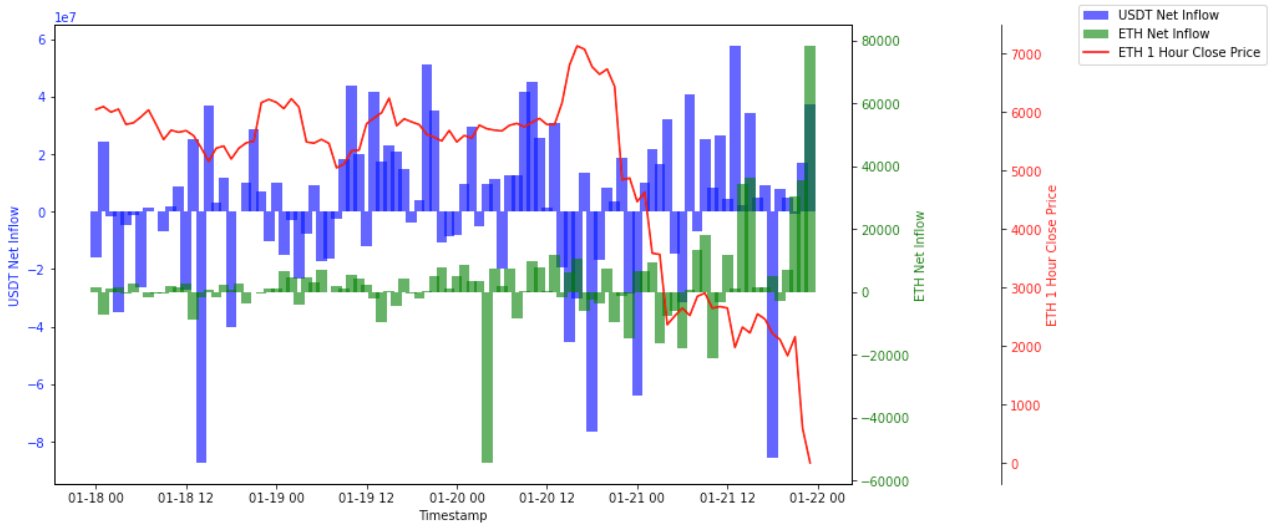

---

[8] Our data sources are BiBox, BigONE, Binance, Bitfinex, Bitget, Bithumb, BitMEX, Bitstamp, Bittrex, Bybit, CheckSig, Cobinhood, Coinbase, Coincheck, CoinEx, Crypto.com, Deribit, Gate.io, Gemini, HitBTC, Huobi, Korbit, Kraken, KuCoin, Luno, Nexo, OKX, Poloniex, SwissBorg, and ZB.com.

**Figure 1.c.** USDT and BTC Net Inflows to Forecast BTC Returns in 1 Hour from November 6 to November 10, 2022

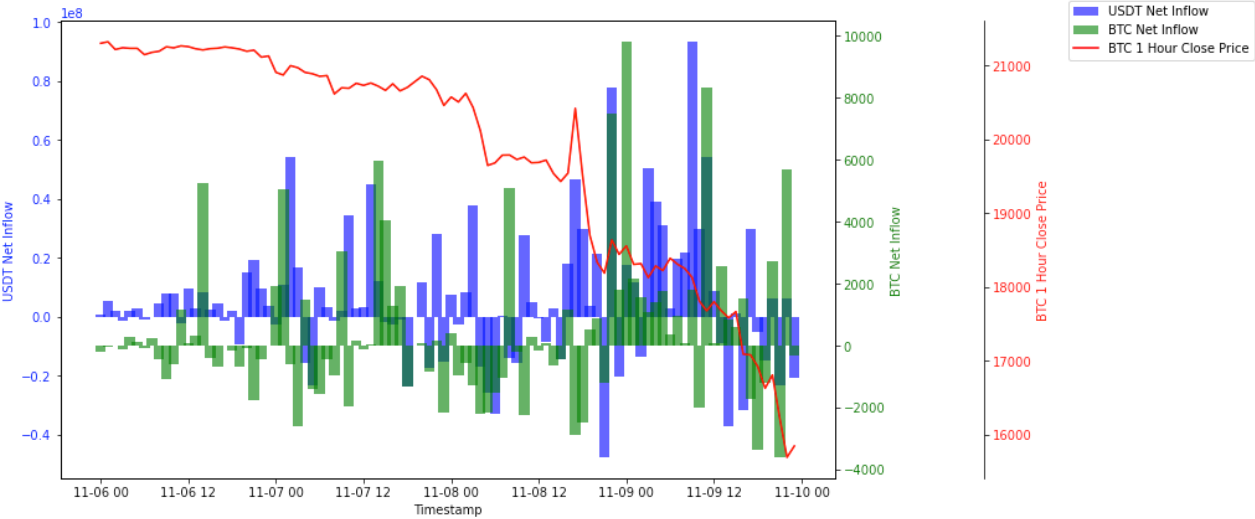

**Figure 1.d.** USDT and BTC Net Inflows to Forecast BTC Returns in 1 Hour from May 9 to May 14, 2022

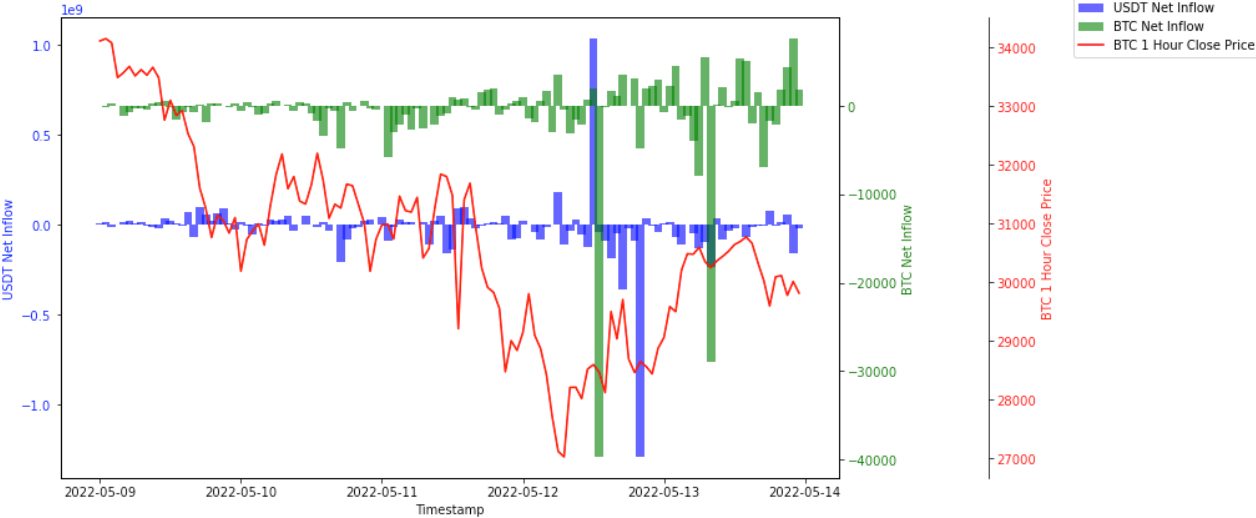

# Table 1
# Summary Heatmaps for Key Empirical Contributions

Table 1 presents summary heatmaps for key empirical contributions on statistically significant relationships between USDT, ETH, and BTC net inflows and ETH and BTC returns and volatility. Our data sources are a list of exchanges[9]. We indicate green and red represents for positive and negative statistically significant relationships respectively. For no colors, we indicate a lack of statistically significant relationship. We have indicated $\beta_1$ coefficients on USDT and ETH net inflows for single- and double-variable regression models to predict the next 1-, 2-, 3-, 4-, and 6-hour ETH returns and volatility. Our sample period is from December 16, 2017 to January 20, 2023. We use ***, **, and * to denote significance at the 1%, 5%, and 10% levels (two-sided), correspondingly.

**Panel A. Forecasting Power of USDT and ETH on ETH Returns and Volatility**

| | Variables for Regression Models | USDT Net Inflows | | ETH Net Inflows | |
|---|---|---|---|---|---|
| | | ETH Return | ETH Volatility | ETH Return | ETH Volatility |
| 1-hour | Single | $1.1\times10^{-5}$*** | $-2.0\times10^{-4}$* | $-0.017$*** | $-0.80$*** |
| | Double | $1.1\times10^{-5}$*** | $5.9\times10^{-5}$ | $-0.017$*** | $-0.17$ |
| 2-hour | Single | $7.5\times10^{-6}$** | $-1.9\times10^{-4}$ | $-0.0078$** | $-0.88$*** |
| | Double | $8.4\times10^{-6}$*** | $-3.5\times10^{-5}$ | $-0.0083$** | $-0.37$*** |
| 3-hour | Single | $1.5\times10^{-6}$ | $-1.5\times10^{-4}$ | $-0.0088$* | $-0.89$*** |
| | Double | $2.5\times10^{-6}$ | $-1.9\times10^{-5}$ | $-0.010$** | $-0.35$*** |
| 4-hour | Single | $-8.3\times10^{-8}$ | $-1.2\times10^{-4}$ | $-0.013$** | $-0.87$*** |
| | Double | $7.1\times10^{-7}$ | $3.3\times10^{-5}$ | $-0.014$** | $-0.22$** |
| 6-hour | Single | $-5.4\times10^{-7}$ | $-2.9\times10^{-4}$** | $-0.026$*** | $-0.87$*** |
| | Double | $1.3\times10^{-7}$ | $-1.9\times10^{-4}$** | $-0.027$*** | $-0.30$** |

**Panel B. Forecasting Power of USDT and BTC on BTC Returns and Volatility**

| | Variables for Regression Models | USDT Net Inflows | | BTC Net Inflows | |
|---|---|---|---|---|---|
| | | BTC Return | BTC Volatility | BTC Return | BTC Volatility |
| 1-hour | Single | $6.3\times10^{-6}$*** | $-1.1\times10^{-4}$ | $0.0053$ | $-17$*** |
| | Double | $6.5\times10^{-6}$*** | $1.8\times10^{-5}$ | $0.0092$ | $-0.99$ |
| 2-hour | Single | $4.3\times10^{-6}$* | $-1.1\times10^{-4}$ | $0.014$ | $-13$*** |
| | Double | $4.8\times10^{-6}$** | $-4.7\times10^{-5}$ | $0.018$ | $0.17$ |
| 3-hour | Single | $-7.6\times10^{-7}$ | $-5.1\times10^{-5}$ | $0.024$ | $-13$*** |
| | Double | $-2.1\times10^{-8}$ | $7.0\times10^{-6}$ | $0.028$ | $-2.8$*** |
| 4-hour | Single | $0.0$ | $-4.6\times10^{-5}$ | $0.095$** | $-11$*** |
| | Double | $-1.7\times10^{-6}$ | $1.8\times10^{-5}$ | $0.099$** | $-0.20$ |
| 6-hour | Single | $-2.0\times10^{-6}$ | $-2.3\times10^{-4}$** | $0.073$ | $-17$*** |
| | Double | $-1.7\times10^{-6}$ | $-1.9\times10^{-4}$*** | $0.076$ | $-4.2$*** |

---

[9] Our data sources are BiBox, BigONE, Binance, Bitfinex, Bitget, Bithumb, BitMEX, Bitstamp, Bittrex, Bybit, CheckSig, Cobinhood, Coinbase, Coincheck, CoinEx, Crypto.com, Deribit, Gate.io, Gemini, HitBTC, Huobi, Korbit, Kraken, KuCoin, Luno, Nexo, OKX, Poloniex, SwissBorg, and ZB.com.

## Table 2

## Regression Results for ETH Return-Forecasting

Table 2 presents regression results on the return-forecasting power of net inflows. 'USDT v BTC' and 'BTC v BTC' use USDT and BTC net inflows correspondingly to predict BTC returns. Our data sources are the same as those in Table 1. The first and second rows for each pair represent single- and double-variable models correspondingly. Returns refer to returns of ETH and net inflows are calculated as the difference between inflows (from wallets to exchanges) and outflows (from exchanges to wallets). $\beta_0$ is intercept; $\beta_1$ is coefficient on net inflow; $\beta_2$ is control variable (return at t); standard errors in parentheses. $\beta_1$ is key for interpretations. The forecasting frequencies are the next 1, 2, 3, 4, and 6 hours. Our sample period is from December 16, 2017 to January 20, 2023. We use ***, **, and * to denote significance at the 1%, 5%, and 10% levels (two-sided), correspondingly.

**Panel A. USDT Net Inflows on ETH Returns**

| Variable | $\beta_0$ | $\beta_1$ | $\beta_2$ | $R^2$ Adjusted |
|---|---|---|---|---|
| 1-hour return - USDT v ETH | $8.2\times10^{-5}$ | $1.1\times10^{-5}$*** | | 0.001 |
| | (1.600) | (5.903) | | |
| | $8.5\times10^{-5}$* | $1.1\times10^{-5}$*** | -0.030*** | 0.002 |
| | (1.650) | (5.978) | (-6.407) | |
| 2-hour return - USDT v ETH | $2.0\times10^{-4}$ | $7.5\times10^{-6}$*** | | 0.000 |
| | (1.629) | (2.595) | | |
| | $2.0\times10^{-4}$* | $8.4\times10^{-6}$*** | -0.056*** | 0.003 |
| | (1.711) | (2.900) | (-8.481) | |
| 3-hour return - USDT v ETH | $3.0\times10^{-4}$* | $1.5\times10^{-6}$ | | -0.000 |
| | (1.697) | (0.444) | | |
| | $3.0\times10^{-4}$* | $2.5\times10^{-6}$ | -0.0562*** | 0.003 |
| | (1.774) | (0.733) | (-6.898) | |
| 4-hour return - USDT v ETH | $3.0\times10^{-4}$ | $-8.3\times10^{-8}$ | | -0.000 |
| | (1.644) | (-0.021) | | |
| | $3.0\times10^{-4}$* | $7.1\times10^{-7}$ | -0.036*** | 0.001 |
| | (1.694) | (0.176) | (-3.835) | |
| 6-hour return - USDT v ETH | $5.0\times10^{-4}$ | $-5.4\times10^{-7}$ | | -0.000 |
| | (1.528) | (-0.114) | | |
| | $5.0\times10^{-4}$ | $1.3\times10^{-7}$ | -0.038*** | 0.001 |
| | (1.582) | (0.027) | (-3.276) | |

**Panel B. ETH Net Inflows on ETH Returns**

| Variable | $\beta_0$ | $\beta_1$ | $\beta_2$ | $R^2$ Adjusted |
|---|---|---|---|---|
| 1-hour return - ETH v ETH | 0.000093* | -0.017*** | | 0.001 |
| | (1.829) | (-5.809) | | |
| | 0.000096* | -0.017*** | -0.031*** | 0.002 |
| | (1.889) | (-5.958) | (-6.727) | |
| 2-hour return - ETH v ETH | 0.0002* | -0.0078** | | 0.000 |
| | (1.835) | (-1.986) | | |
| | 0.0002* | -0.0083** | -0.038*** | 0.002 |
| | (1.904) | (-2.137) | (-5.822) | |
| 3-hour return - ETH v ETH | 0.0003* | -0.0088* | | 0.000 |
| | (1.824) | (-1.837) | | |
| | 0.0003* | -0.010** | -0.056*** | 0.003 |
| | (1.920) | (-2.089) | (-6.964) | |
| 4-hour return - ETH v ETH | 0.0003* | -0.013** | | 0.000 |
| | (1.694) | (-2.491) | | |
| | 0.0003* | -0.014** | -0.030*** | 0.001 |
| | (1.746) | (-2.577) | (-3.252) | |
| 6-hour return - ETH v ETH | 0.0005* | -0.026*** | | 0.002 |
| | (1.810) | (-3.750) | | |
| | 0.0005* | -0.027*** | -0.025** | 0.002 |
| | (1.851) | (-3.804) | (-2.218) | |

# Table 3

# Regression Results for BTC Return-Forecasting

Table 3 presents regression results on the return-forecasting power of net inflows, using the same data sources as Table 1. 'USDT v BTC' and 'BTC v BTC' use USDT and BTC net inflows correspondingly to predict BTC returns. The first and second rows for each pair represent single- and double-variable models correspondingly. Returns refer to returns of BTC and net inflows are calculated as the difference between inflows (from wallets to exchanges) and outflows (from exchanges to wallets). $\beta_0$ is intercept; $\beta_1$ is coefficient on net inflow; $\beta_2$ is control variable (return at t); standard errors in parentheses. $\beta_1$ is key for interpretations. The forecasting frequencies are the next 1, 2, 3, 4, and 6 hours. Our sample period is from December 16, 2017 to January 20, 2023. We use ***, **, and * to denote significance at the 1%, 5%, and 10% levels (two-sided), correspondingly.

**Panel A. USDT Net Inflows on BTC Returns**

| Variable | $\beta_0$ | $\beta_1$ | $\beta_2$ | $R^2$ Adjusted |
|---|---|---|---|---|
| 1-hour return - USDT v BTC | 0.000061 | $6.3 \times 10^{-6}$*** | | 0.000 |
| | (1.488) | (4.196) | | |
| | 0.000063 | $6.5 \times 10^{-6}$*** | -0.0371*** | 0.002 |
| | (1.538) | (4.306) | (-7.901) | |
| 2-hour return - USDT v BTC | 0.0001 | $4.3 \times 10^{-6}$* | | 0.000 |
| | (1.409) | (1.872) | | |
| | 0.0001 | $4.8 \times 10^{-6}$** | -0.0433*** | 0.002 |
| | (1.462) | (2.105) | (-6.502) | |
| 3-hour return - USDT v BTC | 0.00 | $-7.6 \times 10^{-7}$ | | -0.000 |
| | (1.529) | (-0.279) | | |
| | 0.0002 | $-2.1 \times 10^{-8}$ | -0.052*** | 0.003 |
| | (1.579) | (-0.008) | (-6.324) | |
| 4-hour return - USDT v BTC | 0.0002 | 0.0 | | -0.000 |
| | (1.608) | (-0.720) | | |
| | 0.0003 | $-1.7 \times 10^{-6}$ | -0.034*** | 0.001 |
| | (1.644) | (-0.536) | (-3.578) | |
| 6-hour return - USDT v BTC | 0.0004 | $-2.0 \times 10^{-6}$ | | -0.000 |
| | (1.547) | (-0.490) | | |
| | 0.00040 | $-1.7 \times 10^{-6}$ | -0.018 | 0.000 |
| | (1.564) | (-0.408) | (-1.537) | |

**Panel B. BTC Net Inflows on BTC Returns**

| Variable | $\beta_0$ | $\beta_1$ | $\beta_2$ | $R^2$ Adjusted |
|---|---|---|---|---|
| 1-hour return - BTC v BTC | 0.000051 | 0.0053 | | -0.000 |
| | (1.254) | (0.208) | | |
| | 0.000052 | 0.0092 | -0.039*** | 0.001 |
| | (1.296) | (0.357) | (-8.168) | |
| 2-hour return - BTC v BTC | 0.000091 | 0.014 | | -0.000 |
| | (1.155) | (0.389) | | |
| | 0.000094 | 0.018 | -0.041*** | 0.002 |
| | (1.197) | (0.522) | (-6.145) | |
| 3-hour return - BTC v BTC | 0.0001 | 0.024 | | -0.000 |
| | (1.224) | (0.585) | | |
| | 0.0001 | 0.028 | -0.025*** | 0.001 |
| | (1.246) | (0.666) | (-3.055) | |
| 4-hour return - BTC v BTC | 0.0002 | 0.095** | | 0.000 |
| | (1.268) | (2.022) | | |
| | 0.0002 | 0.099** | -0.031*** | 0.001 |
| | (1.298) | (2.106) | (-3.221) | |
| 6-hour return - BTC v BTC | 0.0003 | 0.073 | | 0.000 |
| | (1.143) | (1.304) | | |
| | 0.0003 | 0.076 | -0.0089 | 0.000 |
| | (1.152) | (1.340) | (-0.762) | |

# Table 4

## Regression Results for USDT v ETH and ETH v ETH Volatility-Forecasting

Table 4 presents regression results on the volatility-forecasting power of net inflows, using the same data sources as Table 1. 'USDT v ETH' and 'ETH v ETH' use USDT and ETH net inflows correspondingly to predict the volatility of ETH. For each pair, the first and second rows show single- and two-variable models, respectively. Volatility refers to the volatility of ETH, and net inflows are inflows (wallets to exchanges) minus outflows (exchanges to wallets). $\beta_0$ is the intercept, $\beta_1$ is the net inflow coefficient (of primary interest), and $\beta_2$ is the control variable coefficient (ETH return at time t); standard errors are in parentheses. The sample spans December 16, 2017 to January 20, 2023. We use ***, **, and * to denote significance at the 1%, 5%, and 10% levels (two-sided), correspondingly.

**Panel A. USDT Net Inflows on the Volatility of ETH**

| Variable | $\beta_0$ | $\beta_1$ | $\beta_2$ | $R^2$ Adjusted |
|---|---|---|---|---|
| 1-hour volatility - USDT v ETH | 0.82*** | $-2.0 \times 10^{-4}$* | | 0.000 |
| | (272.257) | (-1.854) | | |
| | 0.25*** | $5.9 \times 10^{-5}$ | 0.69*** | 0.478 |
| | (71.543) | (0.743) | (202.376) | |
| 2-hour volatility - USDT v ETH | 0.83*** | $-1.9 \times 10^{-4}$ | | 0.000 |
| | (203.983) | (-1.720) | | |
| | 0.26*** | $-3.5 \times 10^{-5}$ | 0.68*** | 0.468 |
| | (52.273) | (-0.423) | (140.082) | |
| 3-hour volatility - USDT v ETH | 0.84*** | $-1.5 \times 10^{-4}$ | | 0.000 |
| | (172.746) | (-1.342) | | |
| | 0.25*** | $-1.9 \times 10^{-5}$ | 0.70*** | 0.491 |
| | (41.801) | (-0.237) | (119.834) | |
| 4-hour volatility - USDT v ETH | 0.85*** | $-1.2 \times 10^{-4}$ | | 0.000 |
| | (153.324) | (-1.024) | | |
| | 0.26*** | $3.3 \times 10^{-5}$ | 0.70*** | 0.486 |
| | (36.923) | (0.399) | (102.766) | |
| 6-hour volatility - USDT v ETH | 0.86*** | $-2.9 \times 10^{-4}$** | | 0.001 |
| | (130.244) | (-2.528) | | |
| | 0.26*** | $-1.9 \times 10^{-4}$** | 0.70*** | 0.489 |
| | (30.577) | (-2.241) | (84.150) | |

**Panel B. ETH Net Inflows on the Volatility of ETH**

| Variable | $\beta_0$ | $\beta_1$ | $\beta_2$ | $R^2$ Adjusted |
|---|---|---|---|---|
| 1-hour volatility - ETH v ETH | 0.82*** | -0.80*** | | 0.000 |
| | (271.536) | (-4.648) | | |
| | 0.26*** | -0.17 | 0.68*** | 0.462 |
| | (74.094) | (-1.316) | (197.815) | |
| 2-hour volatility - ETH v ETH | 0.83*** | -0.88*** | | 0.001 |
| | (203.998) | (-5.404) | | |
| | 0.28*** | -0.37*** | 0.67*** | 0.448 |
| | (54.535) | (-3.072) | (135.661) | |
| 3-hour volatility - ETH v ETH | 0.84*** | -0.89*** | | 0.002 |
| | (172.963) | (-5.734) | | |
| | 0.27*** | -0.35*** | 0.68*** | 0.469 |
| | (43.469) | (-3.066) | (115.443) | |
| 4-hour volatility - ETH v ETH | 0.85*** | -0.87*** | | 0.003 |
| | (153.836) | (-5.701) | | |
| | 0.27*** | -0.22** | 0.68*** | 0.466 |
| | (38.619) | (-1.969) | (99.185) | |
| 6-hour volatility - ETH v ETH | 0.86*** | -0.87*** | | 0.004 |
| | (130.685) | (-5.431) | | |
| | 0.27*** | -0.30** | 0.68*** | 0.471 |
| | (31.475) | (-2.577) | (81.750) | |

# Table 5

# Regression Results for USDT v BTC and BTC v BTC Volatility-Forecasting

Table 5 presents regression results on the volatility-forecasting power of net inflows, using the same data sources as Table 1. 'USDT v BTC' and 'BTC v BTC' use USDT and BTC net inflows correspondingly to predict the volatility of BTC. For each pair, the first and second rows show single- and two-variable models, respectively. Volatility refers to the volatility of BTC, and net inflows are inflows (wallets to exchanges) minus outflows (exchanges to wallets). $\beta_0$ is the intercept, $\beta_1$ is the net inflow coefficient (of primary interest), and $\beta_2$ is the control variable coefficient (BTC return at time t); standard errors are in parentheses. The sample spans December 16, 2017 to January 20, 2023. We use ***, **, and * to denote significance at the 1%, 5%, and 10% levels (two-sided), correspondingly.

**Panel A. USDT Net Inflows on the Volatility of BTC**

| Variable | $\beta_0$ | $\beta_1$ | $\beta_2$ | $R^2$ Adjusted |
|---|---|---|---|---|
| 1-hour volatility - USDT v BTC | 0.64*** | $-1.1\times10^{-4}$ | | 0.000 |
| | (251.754) | (-1.204) | | |
| | 0.19*** | $1.8\times10^{-5}$ | 0.70*** | 0.490 |
| | (67.922) | (0.271) | (207.147) | |
| 2-hour volatility - USDT v BTC | 0.65*** | $-1.1\times10^{-4}$ | | 0.000 |
| | (188.294) | (-1.182) | | |
| | 0.20*** | $-4.7\times10^{-5}$ | 0.70*** | 0.484 |
| | (49.655) | (-0.686) | (144.655) | |
| 3-hour volatility - USDT v BTC | 0.66*** | $-5.1\times10^{-5}$ | | -0.000 |
| | (159.224) | (-0.534) | | |
| | 0.19*** | $7.0\times10^{-6}$ | 0.71*** | 0.507 |
| | (39.711) | (0.106) | (123.593) | |
| 4-hour volatility - USDT v BTC | 0.67*** | $-4.6\times10^{-5}$ | | -0.000 |
| | (141.350) | (-0.470) | | |
| | 0.20*** | $1.8\times10^{-5}$ | 0.70*** | 0.493 |
| | (35.355) | (0.254) | (104.152) | |
| 6-hour volatility - USDT v BTC | 0.67*** | $-2.3\times10^{-4}$** | | 0.001 |
| | (119.595) | (-2.376) | | |
| | 0.19*** | $-1.9\times10^{-4}$*** | 0.72*** | 0.509 |
| | (28.357) | (-2.791) | (87.696) | |

**Panel B. BTC Net Inflows on the Volatility of BTC**

| Variable | $\beta_0$ | $\beta_1$ | $\beta_2$ | $R^2$ Adjusted |
|---|---|---|---|---|
| 1-hour volatility - BTC v BTC | 0.63*** | -17*** | | 0.003 |
| | (251.900) | (-10.950) | | |
| | 0.19*** | -0.99 | 0.70*** | 0.486 |
| | (68.095) | (-0.869) | (203.892) | |
| 2-hour volatility - BTC v BTC | 0.64*** | -13*** | | 0.003 |
| | (188.469) | (-8.335) | | |
| | 0.19*** | 0.17 | 0.69*** | 0.479 |
| | (49.813) | (0.156) | (141.967) | |
| 3-hour volatility - BTC v BTC | 0.65*** | -13*** | | 0.005 |
| | (159.594) | (-8.985) | | |
| | 0.19*** | -2.8*** | 0.71*** | 0.502 |
| | (40.025) | (-2.700) | (121.143) | |
| 4-hour volatility - BTC v BTC | 0.66*** | -11*** | | 0.006 |
| | (141.786) | (-7.878) | | |
| | 0.20*** | -0.20 | 0.70*** | 0.490 |
| | (35.382) | (-0.198) | (102.365) | |
| 6-hour volatility - BTC v BTC | 0.66*** | -17*** | | 0.015 |
| | (119.748) | (-10.606) | | |
| | 0.19*** | -4.2*** | 0.71*** | 0.504 |
| | (28.626) | (-3.604) | (85.163) | |

## Table 6

## Extreme Events and Option Strategies, Implied Volatilities, Returns, and Related Data

Table 6 shows the event dates, ETH net inflow or outflow, strike price, option maturity, implied volatilities, percentile of implied volatilities, price of option premiums (price * index price), return by buying the call, return by selling the call, delta, underlying return, and portfolio return ETH options return series. Our data sources are the same as those in Table 1. $P_{strike}$, $t_{quote}$, $V$, $P_{option\ premium}$, $R_{buy\ call}$, $R_{sell\ call}$, $\delta$, $R_{ul}$, $R_{port}$ stand for strike price of options, option quote time, implied volatiles of options, price of option premium, return of buying call options, return of selling call options, delta of options at initial timings, returns of underlyings, returns of portfolios. Percentile of implied volatilities is for the implied volatilities among the those in the given sample periods. Portfolio Return = Option Return + Underlying Return and Underlying Return = (Final Index Price – Initial Index Price)/Initial Index Price * Number of Unit of Delta for Initial Index Price. Our sample periods are from November 4, 2021 to May 19, 2022 for BTC short call options and January 1, 2021 to May 19, 2022 for ETH short call options.

| Event Time | Net Inflow/Outflow and Volume | $P_{strike}$ | Maturity | $P_{option\ premium}$ | $\delta$ | $t_{quote}$ | $V$ | $R_{buy\ call}$ | $R_{sell\ call}$ | $R_{ul}$ | $R_{port}$ |
|---|---|---|---|---|---|---|---|---|---|---|---|
| 12:00, May 12, 2022 | Net Inflow, 26,789.01 | 2,000 | 08:00, May 13, 2022 | 42.58 | 0.17 | 13:03, 1 hour later | 187.35 | 2.22% | -2.22% | 0.37% | -1.85% |
| | | | | | | 16:01, 4 hours later | 188.56 | -8.23% | 8.23% | -1.37% | 6.86% |
| | | | | | | 18:00, 6 hours later | 189.82 | -32.04% | 32.04% | -5.34% | 26.70% |
| 14:00, May 11, 2022 | Net Outflow, -36,336.45 | 2,400 | 08:00, May 12, 2022 | 29.18 | 0.20 | 15:13, 1 hour later | 130.52 | -40.74% | 40.74% | -7.99% | -48.73% |
| | | | | | | 15:46, 2 hours later | 131.16 | -60.76% | 60.76% | -11.92% | -72.68% |
| 10:00, July 28, 2021 | Net Inflow, 1,775.31 | 2,400 | 08:00, July 29, 2022 | 13.80 | 0.11 | 11:57, 2 hours later | 102.76 | 43.71% | -43.71% | 5.02% | -38.70% |
| | | | | | | 13:06, 3 hours later | 106.15 | -33.48% | 33.48% | -3.84% | 29.64% |
| 06:00, July 28, 2021 | Net Outflow, -7,816.59 | 2,300 | 08:00, July 28, 2022 | 3.41 | 0.10 | 08:01, 2 hours later | 106.15 | 371.43% | -371.43% | 38.99% | 401.41% |
| | | | | | | 10:06, 4 hours later | 84.31 | 304.32% | -304.32% | 31.94% | 336.26% |
| | | | | | | 11:57, 6 hours later | 86.36 | 481.06% | -481.06% | 50.50% | 531.56% |
| 13:00, January 20, 2021 | Net Inflow, 221,228.69 | 1,280 | 08:00, January 21, 2021 | 32.81 | 0.34 | 16:00, 3 hours later | 175.97 | -1.74% | 1.74% | -0.59% | 1.16% |

## Table 7

## Sell Call Top versus Bottom Net Return Percentile Profits and Losses Comparisons

Table 7 shows profits and losses comparisons for sell ETH call options net return top and bottom percentiles. Our data source is traded data from Deribit. IV, OTM, and WtL in the table stand for Implied Volatilities, Out of the Money, and Win-to-Loss ratios. $R_{avg\ port,\ net}$ and $R_{total\ port,\ net}$ stand for average and total net portfolio returns. $Range_{OTM}$ stands for out-of-the-money range. We present results for top and bottom 10% ETH net inflows, with top and bottom 5% and 1% tested and have results of similar trends. Our sample period is from January 1, 2021 to May 19, 2022.

| | | Top, Sell Call | | | | | Bottom, Sell Call | | | | |
|---|---|---|---|---|---|---|---|---|---|---|---|
| | | Win Rate | Total Trades | WtL | $R_{avg\ port,\ net}$ | $R_{total\ port,\ net}$ | Win Rate | Total Trades | WtL | $R_{avg\ port,\ net}$ | $R_{total\ port,\ net}$ |
| Top and bottom 10% | Original | 54.86% | 381 | 226.02% | 0.38% | 143.94% | 12.34% | 381 | 119.12% | -0.80% | -303.47% |
| | IV ≥ 1 | 57.51% | 233 | 280.36% | 0.56% | 130.03% | 9.77% | 256 | 140.85% | -0.96% | -246.25% |
| | IV ≥ 2 | 73.91% | 46 | 290.30% | 1.41% | 64.76% | 17.78% | 45 | 143.32% | -1.18% | -53.29% |
| | $Range_{OTM}$ < 1% | 56.16% | 73 | 144.97% | 0.21% | 15.23% | 9.52% | 84 | 178.69% | -0.46% | -38.75% |
| | 1% ≤ $Range_{OTM}$ < 3% | 59.68% | 124 | 220.12% | 0.50% | 61.94% | 12.93% | 116 | 75.98% | -0.93% | -107.44% |
| | 3% ≤ $Range_{OTM}$ < 5% | 63.93% | 61 | 226.46% | 0.63% | 38.31% | 17.91% | 67 | 60.91% | -0.79% | -52.72% |
| | 5% ≤ $Range_{OTM}$ < 10% | 50.63% | 79 | 247.96% | 0.33% | 25.84% | 9.76% | 82 | 195.45% | -1.03% | -84.52% |
| | IV ≥ 1, $Range_{OTM}$ < 1% | 65.71% | 35 | 212.80% | 0.14% | 113.57% | 5.41% | 37 | 208.05% | -0.53% | -223.56% |
| | IV ≥ 1, 1% ≤ $Range_{OTM}$ < 3% | 60.32% | 35 | 271.01% | 0.75% | 47.08% | 7.46% | 37 | 80.53% | -1.20% | -80.17% |
| | IV ≥ 1, 3% ≤ $Range_{OTM}$ < 5% | 70.37% | 98 | 299.26% | 1.26% | 34.07% | 15.56% | 104 | 62.38% | -0.93% | -41.65% |
| | IV ≥ 1, 5% ≤ $Range_{OTM}$ < 10% | 60.94% | 125 | 214.08% | 0.47% | 29.80% | 9.33% | 149 | 209.82% | -1.09% | -81.70% |
| | IV ≥ 2, $Range_{OTM}$ < 1% | 100.00% | 3 | N/A | 2.17% | 6.52% | 0.00% | 5 | N/A | -0.77% | -3.85% |
| | IV ≥ 2, 1% ≤ $Range_{OTM}$ < 3% | 92.31% | 3 | 405.66% | 2.17% | 28.16% | 14.29% | 5 | 34.37% | -1.86% | -13.01% |
| | IV ≥ 2, 3% ≤ $Range_{OTM}$ < 5% | 100.00% | 16 | N/A | 3.94% | 15.76% | 42.86% | 12 | 73.91% | -0.44% | -3.10% |
| | IV ≥ 2, 5% ≤ $Range_{OTM}$ < 10% | 75.00% | 20 | 88.51% | 0.82% | 9.84% | 15.38% | 19 | 193.67% | -1.74% | -22.56% |
| Top and bottom 5% | Original | 56.02% | 191 | 282.87% | 0.49% | 93.70% | 11.52% | 191 | 133.27% | -0.80% | -153.06% |
| | IV ≥ 1 | 59.82% | 112 | 414.08% | 0.75% | 84.47% | 7.94% | 126 | 164.82% | -0.99% | -125.29% |
| | IV ≥ 2 | 75.86% | 29 | 608.98% | 1.66% | 48.16% | 17.86% | 28 | 163.94% | -0.96% | -26.77% |
| | $Range_{OTM}$ < 1% | 65.71% | 35 | 116.33% | 0.32% | 11.07% | 7.89% | 38 | 292.86% | -0.43% | -16.25% |
| | 1% ≤ $Range_{OTM}$ < 3% | 59.09% | 66 | 320.67% | 0.66% | 43.84% | 9.23% | 65 | 118.33% | -0.93% | -60.34% |
| | 3% ≤ $Range_{OTM}$ < 5% | 60.00% | 30 | 232.14% | 0.52% | 15.74% | 18.18% | 33 | 38.53% | -1.05% | -34.50% |
| | 5% ≤ $Range_{OTM}$ < 10% | 54.29% | 35 | 550.77% | 0.63% | 22.10% | 10.53% | 38 | 170.80% | -0.85% | -32.34% |
| | IV ≥ 1, $Range_{OTM}$ < 1% | 80.00% | 15 | 194.93% | 0.16% | 73.41% | 5.26% | 19 | 434.56% | -0.46% | -115.83% |
| | IV ≥ 1, 1% ≤ $Range_{OTM}$ < 3% | 62.86% | 15 | 398.24% | 1.02% | 35.61% | 2.63% | 19 | 120.58% | -1.23% | -46.65% |
| | IV ≥ 1, 3% ≤ $Range_{OTM}$ < 5% | 60.00% | 50 | 470.03% | 1.26% | 12.61% | 9.52% | 57 | 33.80% | -1.43% | -30.03% |
| | IV ≥ 1, 5% ≤ $Range_{OTM}$ < 10% | 70.37% | 60 | 593.86% | 0.90% | 24.24% | 9.68% | 78 | 198.93% | -0.95% | -29.52% |
| | IV ≥ 2, $Range_{OTM}$ < 1% | 100.00% | 3 | N/A | 2.17% | 6.52% | 0.00% | 4 | N/A | -0.69% | -2.75% |
| | IV ≥ 2, 1% ≤ $Range_{OTM}$ < 3% | 90.00% | 3 | 411.99% | 2.13% | 21.31% | 0.00% | 4 | N/A | -2.09% | -10.46% |
| | IV ≥ 2, 3% ≤ $Range_{OTM}$ < 5% | 100.00% | 13 | N/A | 3.68% | 11.03% | 40.00% | 9 | 25.47% | -1.08% | -5.42% |
| | IV ≥ 2, 5% ≤ $Range_{OTM}$ < 10% | 100.00% | 16 | N/A | 2.58% | 7.75% | 16.67% | 14 | 288.74% | -0.82% | -4.90% |
| Top and bottom 1% | Original | 58.97% | 39 | 453.42% | 0.87% | 33.78% | 15.38% | 39 | 132.99% | -0.72% | -28.26% |
| | IV ≥ 1 | 65.22% | 23 | 672.70% | 1.29% | 29.61% | 11.11% | 27 | 197.68% | -0.89% | -24.06% |
| | IV ≥ 2 | 80.00% | 10 | 1066.29% | 2.22% | 22.21% | 18.18% | 11 | 287.17% | -0.41% | -4.46% |
| | $Range_{OTM}$ < 1% | 42.86% | 7 | 176.99% | 0.12% | 0.86% | 12.50% | 8 | 158.55% | -0.39% | -3.11% |
| | 1% ≤ $Range_{OTM}$ < 3% | 70.59% | 17 | 431.99% | 1.13% | 19.16% | 0.00% | 9 | N/A | -1.35% | -12.16% |
| | 3% ≤ $Range_{OTM}$ < 5% | 66.67% | 6 | 2051.00% | 1.99% | 11.92% | 30.00% | 10 | 38.45% | -0.66% | -6.64% |
| | 5% ≤ $Range_{OTM}$ < 10% | 66.67% | 3 | 371.27% | 0.68% | 2.03% | 12.50% | 8 | 2.47% | -1.03% | -8.20% |
| | IV ≥ 1, $Range_{OTM}$ < 1% | 50.00% | 2 | 153.09% | 0.33% | 29.23% | 0.00% | 4 | N/A | -0.46% | -21.47% |
| | IV ≥ 1, 1% ≤ $Range_{OTM}$ < 3% | 77.78% | 2 | 550.84% | 1.80% | 16.17% | 0.00% | 4 | N/A | -1.62% | -9.71% |
| | IV ≥ 1, 3% ≤ $Range_{OTM}$ < 5% | 75.00% | 11 | 2850.70% | 2.73% | 10.90% | 20.00% | 10 | 67.76% | -1.08% | -5.40% |
| | IV ≥ 1, 5% ≤ $Range_{OTM}$ < 10% | 100.00% | 15 | N/A | 1.18% | 2.35% | 12.50% | 15 | 2.47% | -1.03% | -8.20% |
| | IV ≥ 2, $Range_{OTM}$ < 1% | N/A | 0 | N/A | N/A | 0.00% | 0.00% | 3 | N/A | -0.79% | -2.37% |
| | IV ≥ 2, 1% ≤ $Range_{OTM}$ < 3% | 100.00% | 0 | N/A | 2.87% | 11.48% | N/A | 3 | N/A | N/A | 0.00% |
| | IV ≥ 2, 3% ≤ $Range_{OTM}$ < 5% | 100.00% | 4 | N/A | 3.68% | 11.03% | 50.00% | 3 | 36.68% | -0.95% | -1.90% |
| | IV ≥ 2, 5% ≤ $Range_{OTM}$ < 10% | N/A | 7 | N/A | N/A | 0.00% | 0.00% | 5 | N/A | -1.02% | -2.03% |

# Table 8

**Sell Call Top versus Bottom Buy Call Net Return Percentile Profits and Losses Comparisons**

Table 8 shows profits and losses comparisons for sell ETH call options net return top and buy ETH call options bottom percentiles. Our data source is traded data from Deribit. IV, OTM, and WtL in the table stand for Implied Volatilities, Out of the Money, and Win-to-Loss ratios. $R_{avg\ port,\ net}$ and $R_{total\ port,\ net}$ stand for average and total net portfolio returns. $Range_{OTM}$ stands for out-of-the-money range. We present results for top and bottom 10% ETH net inflows, with top and bottom 5% and 1% tested and have results of similar trends. Our sample period is from January 1, 2021 to May 19, 2022.

| | | Top, Sell Call | | | | | Bottom, Buy Call | | | | |
|---|---|---|---|---|---|---|---|---|---|---|---|
| | | Win Rate | Total Trades | WtL | $R_{avg\ port,\ net}$ | $R_{total\ port,\ net}$ | Win Rate | Total Trades | WtL | $R_{avg\ port,\ net}$ | $R_{total\ port,\ net}$ |
| Top and bottom 10% | Original | 54.86% | 381 | 226.02% | 0.38% | 143.94% | 12.34% | 381 | 115.86% | -1.18% | -448.52% |
| | IV ≥ 1 | 57.51% | 233 | 280.36% | 0.56% | 130.03% | 9.77% | 256 | 123.39% | -1.35% | -345.82% |
| | IV ≥ 2 | 73.91% | 46 | 290.30% | 1.41% | 64.76% | 17.78% | 45 | 116.97% | -1.53% | -68.90% |
| | $Range_{OTM}$ < 1% | 56.16% | 73 | 144.97% | 0.21% | 15.23% | 9.52% | 84 | 185.31% | -1.04% | -87.51% |
| | 1% ≤ $Range_{OTM}$ < 3% | 59.68% | 124 | 220.12% | 0.50% | 61.94% | 12.93% | 116 | 82.83% | -1.58% | -183.82% |
| | 3% ≤ $Range_{OTM}$ < 5% | 63.93% | 61 | 226.46% | 0.63% | 38.31% | 17.91% | 67 | 56.71% | -1.03% | -69.20% |
| | 5% ≤ $Range_{OTM}$ < 10% | 50.63% | 79 | 247.96% | 0.33% | 25.84% | 9.76% | 82 | 192.94% | -1.08% | -88.16% |
| | IV ≥ 1, $Range_{OTM}$ < 1% | 65.71% | 35 | 212.80% | 0.14% | 113.57% | 5.41% | 37 | 205.19% | -0.85% | -294.79% |
| | IV ≥ 1, 1% ≤ $Range_{OTM}$ < 3% | 60.32% | 35 | 271.01% | 0.75% | 47.08% | 7.46% | 37 | 85.08% | -2.04% | -136.59% |
| | IV ≥ 1, 3% ≤ $Range_{OTM}$ < 5% | 70.37% | 98 | 299.26% | 1.26% | 34.07% | 15.56% | 104 | 54.70% | -1.19% | -53.49% |
| | IV ≥ 1, 5% ≤ $Range_{OTM}$ < 10% | 60.94% | 125 | 214.08% | 0.47% | 29.80% | 9.33% | 149 | 207.77% | -1.13% | -84.88% |
| | IV ≥ 2, $Range_{OTM}$ < 1% | 100.00% | 3 | N/A | 2.17% | 6.52% | 0.00% | 5 | N/A | -1.63% | -8.13% |
| | IV ≥ 2, 1% ≤ $Range_{OTM}$ < 3% | 92.31% | 3 | 405.66% | 2.17% | 28.16% | 14.29% | 5 | 20.50% | -3.36% | -23.53% |
| | IV ≥ 2, 3% ≤ $Range_{OTM}$ < 5% | 100.00% | 16 | N/A | 3.94% | 15.76% | 42.86% | 12 | 64.06% | -0.62% | -4.37% |
| | IV ≥ 2, 5% ≤ $Range_{OTM}$ < 10% | 75.00% | 20 | 88.51% | 0.82% | 9.84% | 15.38% | 19 | 190.72% | -1.71% | -22.24% |
| Top and bottom 5% | Original | 56.02% | 191 | 282.87% | 0.49% | 93.70% | 11.52% | 191 | 133.75% | -1.20% | -229.80% |
| | IV ≥ 1 | 59.82% | 112 | 414.08% | 0.75% | 84.47% | 7.94% | 126 | 144.26% | -1.44% | -181.97% |
| | IV ≥ 2 | 75.86% | 29 | 608.98% | 1.66% | 48.16% | 17.86% | 28 | 122.66% | -1.41% | -39.54% |
| | $Range_{OTM}$ < 1% | 65.71% | 35 | 116.33% | 0.32% | 11.07% | 7.89% | 38 | 295.81% | -0.97% | -36.85% |
| | 1% ≤ $Range_{OTM}$ < 3% | 59.09% | 66 | 320.67% | 0.66% | 43.84% | 9.23% | 65 | 137.89% | -1.59% | -103.27% |
| | 3% ≤ $Range_{OTM}$ < 5% | 60.00% | 30 | 232.14% | 0.52% | 15.74% | 18.18% | 33 | 39.10% | -1.38% | -45.64% |
| | 5% ≤ $Range_{OTM}$ < 10% | 54.29% | 35 | 550.77% | 0.63% | 22.10% | 10.53% | 38 | 161.16% | -0.91% | -34.42% |
| | IV ≥ 1, $Range_{OTM}$ < 1% | 80.00% | 15 | 194.93% | 0.16% | 73.41% | 5.26% | 19 | 427.58% | -0.70% | -160.54% |
| | IV ≥ 1, 1% ≤ $Range_{OTM}$ < 3% | 62.86% | 15 | 398.24% | 1.02% | 35.61% | 2.63% | 19 | 140.94% | -2.13% | -80.89% |
| | IV ≥ 1, 3% ≤ $Range_{OTM}$ < 5% | 60.00% | 50 | 470.03% | 1.26% | 12.61% | 9.52% | 57 | 32.14% | -1.85% | -38.89% |
| | IV ≥ 1, 5% ≤ $Range_{OTM}$ < 10% | 70.37% | 60 | 593.86% | 0.90% | 24.24% | 9.68% | 78 | 189.08% | -1.00% | -31.14% |
| | IV ≥ 2, $Range_{OTM}$ < 1% | 100.00% | 3 | N/A | 2.17% | 6.52% | 0.00% | 4 | N/A | -1.47% | -5.86% |
| | IV ≥ 2, 1% ≤ $Range_{OTM}$ < 3% | 90.00% | 3 | 411.99% | 2.13% | 21.31% | 0.00% | 4 | N/A | -3.73% | -18.67% |
| | IV ≥ 2, 3% ≤ $Range_{OTM}$ < 5% | 100.00% | 13 | N/A | 3.68% | 11.03% | 40.00% | 9 | 25.62% | -1.32% | -6.61% |
| | IV ≥ 2, 5% ≤ $Range_{OTM}$ < 10% | 100.00% | 16 | N/A | 2.58% | 7.75% | 16.67% | 14 | 278.06% | -0.85% | -5.08% |
| Top and bottom 1% | Original | 58.97% | 39 | 453.42% | 0.87% | 33.78% | 15.38% | 39 | 132.99% | -0.72% | -28.26% |
| | IV ≥ 1 | 65.22% | 23 | 672.70% | 1.29% | 29.61% | 11.11% | 27 | 197.68% | -0.89% | -24.06% |
| | IV ≥ 2 | 80.00% | 10 | 1066.29% | 2.22% | 22.21% | 18.18% | 11 | 287.17% | -0.41% | -4.46% |
| | $Range_{OTM}$ < 1% | 42.86% | 7 | 176.99% | 0.12% | 0.86% | 12.50% | 8 | 158.55% | -0.39% | -3.11% |
| | 1% ≤ $Range_{OTM}$ < 3% | 70.59% | 17 | 431.99% | 1.13% | 19.16% | 0.00% | 9 | N/A | -1.35% | -12.16% |
| | 3% ≤ $Range_{OTM}$ < 5% | 66.67% | 6 | 2051.00% | 1.99% | 11.92% | 30.00% | 10 | 38.45% | -0.66% | -6.64% |
| | 5% ≤ $Range_{OTM}$ < 10% | 66.67% | 3 | 371.27% | 0.68% | 2.03% | 12.50% | 8 | 2.47% | -1.03% | -8.20% |
| | IV ≥ 1, $Range_{OTM}$ < 1% | 50.00% | 2 | 153.09% | 0.33% | 29.23% | 0.00% | 4 | N/A | -0.46% | -21.47% |
| | IV ≥ 1, 1% ≤ $Range_{OTM}$ < 3% | 77.78% | 2 | 550.84% | 1.80% | 16.17% | 0.00% | 4 | N/A | -1.62% | -9.71% |
| | IV ≥ 1, 3% ≤ $Range_{OTM}$ < 5% | 75.00% | 11 | 2850.70% | 2.73% | 10.90% | 20.00% | 10 | 67.76% | -1.08% | -5.40% |
| | IV ≥ 1, 5% ≤ $Range_{OTM}$ < 10% | 100.00% | 15 | N/A | 1.18% | 2.35% | 12.50% | 15 | 2.47% | -1.03% | -8.20% |
| | IV ≥ 2, $Range_{OTM}$ < 1% | N/A | 0 | N/A | N/A | 0.00% | 0.00% | 3 | N/A | -0.79% | -2.37% |
| | IV ≥ 2, 1% ≤ $Range_{OTM}$ < 3% | 100.00% | 0 | N/A | 2.87% | 11.48% | N/A | 3 | N/A | N/A | 0.00% |
| | IV ≥ 2, 3% ≤ $Range_{OTM}$ < 5% | 100.00% | 4 | N/A | 3.68% | 11.03% | 50.00% | 3 | 36.68% | -0.95% | -1.90% |
| | IV ≥ 2, 5% ≤ $Range_{OTM}$ < 10% | N/A | 7 | N/A | N/A | 0.00% | 0.00% | 5 | N/A | -1.02% | -2.03% |

# Table 9
# Sell Call Breakeven Top Percentiles

Table 9 shows profits and losses comparisons for sell ETH call options breakeven net return top percentiles. Our data source is traded data from Deribit. IV, OTM, and WtL in the table stand for Implied Volatilities, Out of the Money, and Win-to-Loss ratios. $R_{avg\ port,\ net}$ and $R_{total\ port,\ net}$ stand for average and total net portfolio returns. $Range_{OTM}$ stands for out-of-the-money range. We present results for top and bottom 10% ETH net inflows, with top and bottom 5% and 1% tested and have results of similar trends. Our sample period is from January 1, 2021 to May 19, 2022.

|  |  | Top, Breakeven | | | | |
|---|---|---|---|---|---|---|
|  |  | Win Rate | Total Trades | WtL | $R_{avg\ port,\ net}$ | $R_{total\ port,\ net}$ |
| Top 10% | Original | 39.90% | 381 | 150.66% | 0.00% | 0.00% |
|  | IV ≥ 1 | 42.49% | 233 | 181.30% | 0.14% | 33.59% |
|  | IV ≥ 2 | 63.04% | 46 | 215.30% | 0.95% | 43.70% |
|  | $Range_{OTM}$ < 1% | 43.84% | 73 | 120.78% | -0.02% | -1.50% |
|  | 1% ≤ $Range_{OTM}$ < 3% | 47.58% | 124 | 162.84% | 0.18% | 22.46% |
|  | 3% ≤ $Range_{OTM}$ < 5% | 45.90% | 61 | 185.67% | 0.23% | 13.77% |
|  | 5% ≤ $Range_{OTM}$ < 10% | 32.91% | 79 | 135.03% | -0.18% | -14.01% |
|  | IV ≥ 1, $Range_{OTM}$ < 1% | 54.29% | 35 | 166.26% | -0.22% | 25.18% |
|  | IV ≥ 1, 1% ≤ $Range_{OTM}$ < 3% | 49.21% | 35 | 207.87% | 0.41% | 25.75% |
|  | IV ≥ 1, 3% ≤ $Range_{OTM}$ < 5% | 59.26% | 98 | 237.48% | 0.85% | 22.82% |
|  | IV ≥ 1, 5% ≤ $Range_{OTM}$ < 10% | 40.63% | 125 | 133.20% | -0.04% | -2.67% |
|  | IV ≥ 2, $Range_{OTM}$ < 1% | 100.00% | 3 | N/A | 1.94% | 5.82% |
|  | IV ≥ 2, 1% ≤ $Range_{OTM}$ < 3% | 76.92% | 3 | 577.53% | 1.78% | 23.13% |
|  | IV ≥ 2, 3% ≤ $Range_{OTM}$ < 5% | 100.00% | 16 | N/A | 3.53% | 14.11% |
|  | IV ≥ 2, 5% ≤ $Range_{OTM}$ < 10% | 66.67% | 20 | 72.92% | 0.30% | 3.59% |
| Top 5% | Original | 40.31% | 191 | 148.05% | 0.00% | 0.00% |
|  | IV ≥ 1 | 43.75% | 112 | 192.31% | 0.21% | 23.89% |
|  | IV ≥ 2 | 62.07% | 29 | 289.97% | 1.10% | 31.83% |
|  | $Range_{OTM}$ < 1% | 54.29% | 35 | 89.07% | 0.02% | 0.77% |
|  | 1% ≤ $Range_{OTM}$ < 3% | 48.48% | 66 | 173.77% | 0.24% | 16.13% |
|  | 3% ≤ $Range_{OTM}$ < 5% | 36.67% | 30 | 171.72% | 0.00% | -0.09% |
|  | 5% ≤ $Range_{OTM}$ < 10% | 31.43% | 35 | 210.98% | -0.02% | -0.56% |
|  | IV ≥ 1, $Range_{OTM}$ < 1% | 73.33% | 15 | 125.63% | -0.30% | 17.20% |
|  | IV ≥ 1, 1% ≤ $Range_{OTM}$ < 3% | 54.29% | 15 | 211.58% | 0.57% | 19.94% |
|  | IV ≥ 1, 3% ≤ $Range_{OTM}$ < 5% | 40.00% | 50 | 363.37% | 0.68% | 6.82% |
|  | IV ≥ 1, 5% ≤ $Range_{OTM}$ < 10% | 40.74% | 60 | 245.72% | 0.25% | 6.70% |
|  | IV ≥ 2, $Range_{OTM}$ < 1% | 100.00% | 3 | N/A | 1.87% | 5.61% |
|  | IV ≥ 2, 1% ≤ $Range_{OTM}$ < 3% | 70.00% | 3 | 446.35% | 1.63% | 16.29% |
|  | IV ≥ 2, 3% ≤ $Range_{OTM}$ < 5% | 100.00% | 13 | N/A | 3.15% | 9.46% |
|  | IV ≥ 2, 5% ≤ $Range_{OTM}$ < 10% | 100.00% | 16 | N/A | 1.93% | 5.80% |
| Top 1% | Original | 41.03% | 39 | 143.75% | 0.00% | 0.00% |
|  | IV ≥ 1 | 43.48% | 23 | 190.69% | 0.29% | 6.71% |
|  | IV ≥ 2 | 60.00% | 10 | 223.08% | 1.19% | 11.88% |
|  | $Range_{OTM}$ < 1% | 42.86% | 7 | 54.32% | -0.40% | -2.80% |
|  | 1% ≤ $Range_{OTM}$ < 3% | 47.06% | 17 | 203.04% | 0.35% | 6.03% |
|  | 3% ≤ $Range_{OTM}$ < 5% | 66.67% | 6 | 203.55% | 1.05% | 6.28% |
|  | 5% ≤ $Range_{OTM}$ < 10% | 33.33% | 3 | 13.54% | -0.53% | -1.58% |
|  | IV ≥ 1, $Range_{OTM}$ < 1% | 50.00% | 2 | 47.25% | -0.41% | 7.34% |
|  | IV ≥ 1, 1% ≤ $Range_{OTM}$ < 3% | 55.56% | 2 | 248.16% | 0.93% | 8.35% |
|  | IV ≥ 1, 3% ≤ $Range_{OTM}$ < 5% | 75.00% | 11 | 271.14% | 1.77% | 7.09% |
|  | IV ≥ 1, 5% ≤ $Range_{OTM}$ < 10% | 50.00% | 15 | 41.08% | -0.08% | -0.16% |
|  | IV ≥ 2, $Range_{OTM}$ < 1% | N/A | 0 | N/A | N/A | 0.00% |
|  | IV ≥ 2, 1% ≤ $Range_{OTM}$ < 3% | 75.00% | 0 | 331.02% | 1.99% | 7.96% |
|  | IV ≥ 2, 3% ≤ $Range_{OTM}$ < 5% | 100.00% | 4 | N/A | 2.70% | 8.09% |
|  | IV ≥ 2, 5% ≤ $Range_{OTM}$ < 10% | N/A | 7 | N/A | N/A | 0.00% |